\documentclass[twocolumn,showpacs,preprintnumbers,amsmath,amssymb]{revtex4}

\usepackage{graphicx}
\usepackage{dcolumn}
\usepackage{bm}

\begin{document}

\title{Giant Helium Dimers Produced by Photoassociation \\of Ultracold Metastable Atoms }

\author{J. L\'eonard}
\email{leonard@lkb.ens.fr}
\author{M. Walhout}
 \altaffiliation[Permanent address: ]{Calvin College, Grand Rapids, MI, USA. }
\author{A. P. Mosk}
 \altaffiliation[Permanent address: ]{FOM instituut voor plasmafysica
Rijnhuizen, and University of Twente, The Netherlands.}
\author{T. M\"{u}ller}
\altaffiliation[Present address: ]{Institut f\"ur Quantenoptik,
Universit\"at Hannover, Germany.}
\author{M. Leduc}
\author{C. Cohen-Tannoudji}

\affiliation{Ecole Normale Sup\'erieure and Coll\`ege de France\\
Laboratoire Kastler Brossel, 24 rue Lhomond, 75231 Paris Cedex 05,
France}

\date{\today}

\begin{abstract}
We produce giant helium dimers by photoassociation of metastable helium
atoms in a magnetically trapped, ultracold cloud.  The photoassociation
laser is detuned red of the atomic $2^3S_1 - 2^3P_0 $ line and produces
strong heating of the sample when resonant with molecular bound states.
The temperature of the cloud serves as an indicator of the molecular
spectrum. We report good agreement between our spectroscopic
measurements and our calculations of the five bound states belonging to
a $0_u^+$ purely long-range potential well. These previously unobserved
states have classical inner turning points of about 150 $a_0$ and outer
turning points as large as 1150 $a_0$.
\end{abstract}

\pacs{34.20.Cf, 32.80.Pj, 34.50.Gb}
\maketitle

In the purely long-range molecules first proposed by Stwalley {\it et
al.} \cite{Stwalley}, the binding potential depends only on the
long-range part of the atom-atom interaction, and the internuclear
distance is always large compared with ordinary chemical bond lengths.
Theoretical description of these molecules involves only the leading
$C_3/R^3$ terms of the electric dipole-dipole interaction and the fine
structure inside each atom. These well-known interactions allow precise
calculation of potential wells and rovibrational energies. Previous
experimental studies of such spectra in alkali atoms have utilized the
technique of laser-induced photoassociation (PA) in a magneto-optical
trap (MOT) \cite{Lett,Cline}. In addition to testing calculations of
molecular structure, that work has produced precise measurements of
excited-state lifetimes \cite{Mcalexander,Jones,Wang,Gutterres} and has
led to accurate determinations of s-wave scattering lengths for alkali
systems, which are of interest for studies of Bose Einstein condensates
(BECs)\cite{Abraham,Gardner}.

This letter reports novel spectroscopic measurements and calculations
for extraordinarily long-range molecules that are produced when two
$^4$He atoms in the metastable $2^3S_1$ state absorb laser light tuned
close to the $2^3S_1 - 2^3P_0 $ ($D_0$) atomic line at $\lambda=1083$
nm. As compared with the alkali dimers considered previously, the
potential wells are shallower, and molecules are much more tenuous,
with an internuclear distance reaching values as large as $1150 \;a_0$
($a_0\simeq0.53 \AA$, the Bohr radius). At such large distances,
retardation clearly influences the dipole-dipole interaction. Moreover,
these purely long-range molecular states of metastable helium are
distinctive in that each atom carries a high internal energy (the
$2^3S_1$ state lies 20 eV above the ground state). While one normally
expects Penning ionization to destabilize such energetic molecules, we
note that purely long-range interactions might rather suppress this
process, since the atoms are effectively held apart by the same
potential that binds them together. The absence of purely long-range
resonances in recent PA experiments in Utrecht, which monitored ion
production rates in a MOT, is consistent with this reasoning
\cite{PrivateCom}.

\begin{figure}[hbt]
\begin{center}
\scalebox{0.5}{ \centerline{\includegraphics{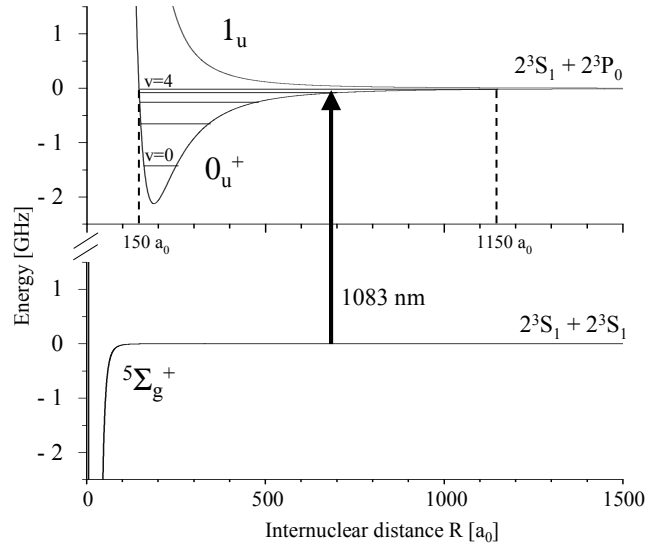}}}
\end{center}
\caption{\footnotesize Molecular potentials involved in the electronic
excitation of two $2^3S_1$ spin-polarized $^4$He atoms. The
$^5\Sigma_g^+$ potential is taken from \cite{Starck}. The excited
potentials are calculated as described in the text and are the only
experimentally accessible potentials linked to the $2^3S_1+2^3P_0$
asymptote. Five bound states (v=0 to 4) lie in the $0_u^+$ well. Their
inner turning points are around $150\; a_0$, and their outer turning
points range from $250\; a_0$ (v=0) to $1150\; a_0$ (v=4).}
\label{Fig:Calculs}
\end{figure}

The present experiment differs significantly from reported PA
measurements with $^4$He in a MOT \cite{Herschbach}. Our atomic cloud
is confined in a magnetic trap and cooled nearly to the BEC transition
\cite{Pereira101,Robert}. The phase space density is typically six
orders of magnitude higher than in a MOT, so the PA process is much
more efficient \cite{Pillet}. In addition, our magnetically trapped
atoms are spin-polarized in a single Zeeman sublevel of the $2^3S_1$
state, so the initial quasi-molecular state is $^5\Sigma_g^+$.
Therefore, only {\it ungerade} excited states are accessible. Finally,
whereas previous PA experiments have monitored only ions or trap
losses, we use absorption imaging to detect heating effects in the
atomic cloud. Molecular resonances that produce no ions or very little
trap loss can still be detected with our method.

Fig.~\ref{Fig:Calculs} shows the only two potentials that can be
excited in our experiment in the vicinity of the $D_0$ atomic line. The
photoassociation experiment consists in driving the transition from
free pairs of $2^3S_1$ atoms to bound states in the purely long-range
$0_u^+$ well connected to the $2^3S_1 + 2^3P_0$ asymptote. Measurements
of the bound-state spectrum proceed as follows. About $10^9$ atoms are
trapped in a MOT before being transferred into an Ioffe-Pritchard
magnetostatic trap. An evaporative cooling sequence that utilizes
RF-induced spin flips cools the atoms to 2-30 $\mu$K. The critical
temperature for Bose-Einstein condensation is in the range 1 to 4
$\mu$K, depending on the density \cite{Pereira101,Pereira2}.

\begin{figure}[htb]
\begin{center}
\scalebox{0.5}{\includegraphics{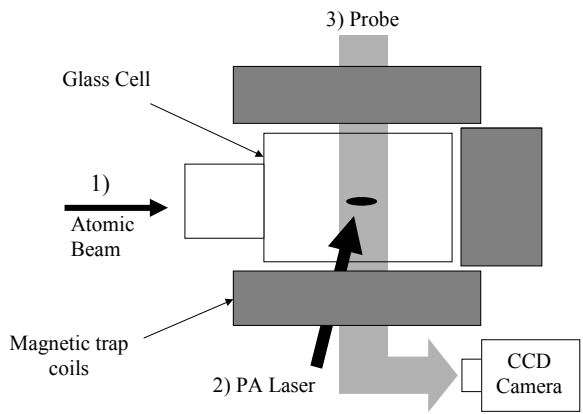}}
\end{center}
\caption{\label{fig:Expe} \footnotesize Experimental setup. 1)
Metastable helium atoms are trapped and evaporated down to a few $\mu$K
and a peak density of order $10^{13}$ cm$^{-3}$. 2) The cloud is
illuminated inside the magnetic trap by the photoassociation (PA) laser
for a few ms. 3) The cloud is released and imaged on a CCD camera after
ballistic expansion.}
\end{figure}

\begin{figure}[t]
\begin{center}
\scalebox{0.55}{\includegraphics{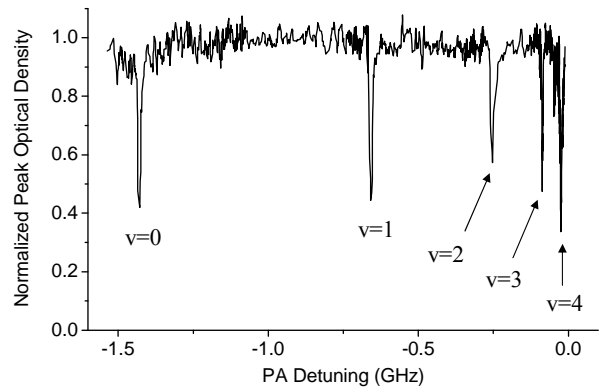}}
\end{center}
\caption{\footnotesize Experimental data. The peak optical density is
plotted versus PA laser detuning, measured with respect to the $2^3S_1$
- $2^3P_0$ atomic transition.  We have subtracted a Lorentzian-shaped
background loss due to atomic absorption and divided by a signal
proportional to the number of atoms in the trap, yielding a
``normalized" peak optical density signal. Each line has been probed
separately with different laser intensities and exposure time ranging
from (0.1 $\mu$W; 20 ms) to (100 $\mu$W; 200 ms). Therefore the
relative depths of the five lines are not relevant. }
\label{Fig:DataWideScan}
\end{figure}

After cooling, the cloud is illuminated for a few ms by light from a
``PA laser" beam containing all polarization components relative to the
magnetic field axis. Just after this PA pulse, the cloud is released
and then detected by means of destructive absorption imaging after
ballistic expansion. The number of atoms, the peak optical density, and
the temperature are thereby measured as functions of the PA laser
frequency. Fig. \ref{Fig:DataWideScan} shows the optical density after
contributions from atomic ($D_0$) absorption have been subtracted. For
this broad scan, the PA laser is a temperature-controlled diode laser
with a 3-MHz spectral width. A Fabry-Perot cavity is used to measure
the frequency relative to a reference laser frequency locked to the
$2^3S_1-2^3P_2$ atomic line. The accuracy of the PA frequency in this
case is 10 MHz.

\begin{figure}[htb]
\begin{center}
\scalebox{0.5}{\includegraphics{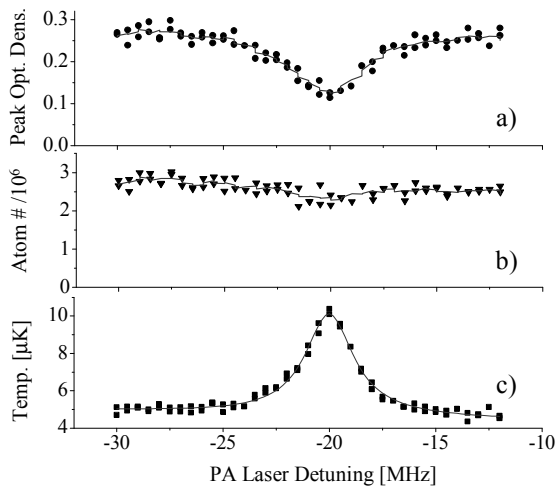}}
\end{center}
\caption{\footnotesize Detection of the line v=4 in the $0_u^+$
potential well. a) Peak optical density, b) atom number and c)
temperature in $\mu$K versus PA laser frequency. Each point represents
a new evaporated cloud after PA pulse illumination and ballistic
expansion. The curves in graphs a) and b) indicate the averaging of
data over 5 adjacent points. The curve in graph c) is a Lorentzian fit
to the data.} \label{Fig:DataCalorimetry}
\end{figure}

More accurate spectroscopy is performed with a PA beam derived from a
cavity-stabilized laser (width 0.3 MHz) locked to the $D_0$ line. The
PA beam is tuned to the molecular lines by acousto-optic modulators.
Its intensity is set between 0.01 and 1 mW/cm$^2$, and the exposure
time between 0.1 and 10 ms. With this technique we can reach and zoom
in on the four highest (v=1 to v=4) lines in the $0_u^+$ well.
Fig.~\ref{Fig:DataCalorimetry} shows typical experimental data for the
line v=4. Figs. 4a) and 4b) demonstrate that the peak optical density
drops significantly when the PA beam is resonant with a molecular line,
even if trap loss is weak. This situation is explained by the strong
heating indicated in Fig. 4c. The temperature increase provides a very
sensitive diagnostic for resonances that produce little loss. Assuming
that the heat deposited in the cloud is proportional to the number of
molecules produced, we fit the frequency dependence of the temperature
to a Lorentzian curve. The FWHM of the molecular lines is measured at
low intensities to be 3.0(3) MHZ, within the expected range for the
molecular radiative width ($\Gamma_{mol} \leqslant 2\Gamma$).

The heating mechanism that produces our signals can be roughly
understood as follows. After absorbing a photon near the outer turning
point, the molecular system can decay radiatively into two fast
metastable atoms, each with a non-zero probability of being in the
trapped state. Since the trap depth is 15 mK, fast atoms can remain
trapped and heat up the whole cloud. This reasoning is supported by the
fact that we can convert the heating into trap loss if, during a 10 ms
thermalization time after the PA pulse, we apply an RF ``knife" that
removes fast atoms from the trap. The simultaneous measurement of small
trap loss and significant temperature increase suggests that each
molecular excitation heats the sample considerably. This is the reason
why the temperature probe is so sensitive.

We measure the detuning $\delta<0$ of the PA laser relative to the
$D_0$ atomic line. The initial $2^3S_1$ pair undergoes an inhomogeneous
Zeeman effect due to the trapping potential. Additionally, the relative
kinetic energy of the pair is characterized by a non-zero temperature
$T$. Each of these effects contributes a $\frac{3}{2}k_BT$ Boltzmann
term to the initial average energy of a pair. Hence, the binding energy
$hb<0$ of a given vibrational state is $hb = h\delta +2\mu B_0+3k_BT$,
where $2\mu B_0>0$ is the Zeeman shift for a $2^3S_1$ pair at the
center of the trap. Additional recoil and mean-field shifts of order 10
kHz are neglected because they are much smaller than our experimental
uncertainty. Each line is probed in different conditions of $T$ (2 - 30
$\mu$K), $B_0$ (300 mG - 10 G) and density ($10^{12}$ - $5\times
10^{13}$ cm$^{-3}$). Within the accuracy of our experiment we finally
find no density or magnetic-field dependence of the binding energies,
and for each line we estimate an uncertainty of $\pm 0.5$ MHz. Given
this uncertainty and the range of magnetic field explored, we can infer
an upper limit for the $0_u^+$ magnetic moment of $\sim 0.04$ Bohr
magnetons \footnote{Since pure $0_u^+$ states are not degenerate, only
molecular rotation gives rise to a magnetic moment, which is therefore
expected to be of the order of the nuclear magneton.}.

In order to interpret the experimental frequency spectrum
quantitatively, we calculate the coupling between the atomic orbitals
of one atom in the $2^{3}S_{1}$ state and another in the
$2^{3}P_{J=0,1,2}$ state, which involves 54 molecular potentials. This
coupling is constructed as a perturbative Hamiltonian in the basis of
fine-structure-free atomic states. At large internuclear distances $R$,
the lowest-order term of the electromagnetic interaction is the
retarded dipole-dipole interaction \cite{Dashevskaya,Meath}, which is
fully determined by the coefficient $C_3=(3/4) \hbar \Gamma /k^3$,
where $\Gamma=2\pi \times 1.6248$ MHz \cite{Drake} is the atomic
linewidth, and $k=2 \pi / \lambda$ is the wave number. This interaction
includes no coupling between electronic orbital and spin angular
momenta ($\vec{L}$ and $\vec{S}$) and is diagonal in the molecular
Hund's case (a) basis.

In the absence of $\vec{L}$-$\vec{S}$ coupling, the potential curves
resulting from dipole-dipole interaction are purely attractive or
repulsive and have a single asymptote. When we include the atomic fine
structure in the Hamiltonian \cite{MovrePichler}, the coupling gives
rise to three distinct asymptotes and to anti-crossings between
attractive and repulsive curves, leading to purely long-range potential
wells \footnote{The absence of hyperfine structure in $^4$He leads to
relatively simple molecular potentials.}. We construct the
fine-structure coupling phenomenologically from $2^3P$ splittings that
have been measured accurately \cite{FineStructSplit1,FineStructSplit2}.
If there is no rotation, only the projection $\Omega$ of the total
electronic angular momentum on the molecular axis remains a good
quantum number. Our experiment probes the $0_u^+$ purely long-range
well plotted in Fig. \ref{Fig:Calculs}.

We also consider how the rotation of the molecule can couple electronic
states ($\Delta \Omega = 0, \pm 1$), an effect that will leave
$\Omega=0$ and $0_u^+$ as only approximate labels. Ultimately, only the
total angular momentum $J$ remains a good quantum number. Our final
addition to the Hamiltonian is the operator $\vec{\ell}\;^2/(2 \mu
R^2)$ \cite{Hougen}, where $\vec{\ell}=\vec{J}-\vec{L}-\vec{S}$ is the
nuclear angular momentum operator. In the s-wave scattering regime, the
initial ($2^3S_1+2^3S_1$) pair exists only in the $J=2$ state.
Elementary group theory and Bose-Einstein statistics for the nuclei
\cite{Hougen} dictate that $J$ in the excited state must be odd, namely
1 or 3. What is more, the Condon radius at which the transition occurs
is so large that the excited molecule is in almost the same quantum
state as a non-interacting pair of atoms in the $2^3S_1+2^3P_0$ state
(Hund's case (c)). Therefore $J=1$ is dominant.

The Hamiltonian including retarded dipole-dipole interaction, fine
structure, and nuclear rotation is diagonalized numerically. The
$0_u^+$ well cited above is found to remain almost pure; that is, the
couplings to other $\Omega$ subspaces are very small. It is 2 GHz deep
and contains 5 bound states (see Fig. \ref{Fig:Calculs}). The inner
turning points of these bound states are around $150 \;a_0$ and the
outer turning point is as large as $1150 \;a_0$ for the v=4 vibrational
state. From the discussion above, it is clear that even the repulsive,
inner part of the potential has purely long-range character; this fact
distinguishes these dimers from those bound by usual chemical
interactions. The large internuclear distance is also the reason why
the next-order term in $C_6/R^6$ of the electromagnetic interaction is
negligible \footnote{We estimate the $C_6/R^6$ correction is at least 4
orders of magnitude smaller than $C_3/R^3$ for an internuclear distance
larger than $150 \;a_0$.}. More detail about the theoretical approach
will be given in a forthcoming paper \footnote{The calculation exhibits
two other ungerade purely long-range wells connected to the
$2^3S_1+2^3P_1$ asymptote. They are even shallower and support fewer
bound states than the well examined here. Their inner turning points
are over 300 $a_0$.}.

\begin{table}[t]
\caption{\footnotesize Measured and calculated binding energies in the
$0_u^+$ well. The first column gives the experimental results.
Vibrational states v=1 to v=4 are measured with a 0.5 MHz uncertainty,
v=0 with a 10 MHz uncertainty. The next two columns give calculations
following the model described in the text for $J=1$ with and without
including retardation effects. The last column contains calculated
$J=3$ results with retardation.}
\begin{ruledtabular}
\begin{tabular}{ccccc}

 & Experiment & $J=1$ & $J=1$\footnote{Calculated while neglecting retardation effects} & $J=3$\\

\hline
v=4 & -18.2(5) & -18.250 & -16.646& ------ \\
v=3 & -79.6(5) & -79.555 & -76.933& -41.761 \\
v=2 & -253.3(5) & -253.27 & -249.47& -175.14 \\
v=1 & -648.5(5) & -650.37 & -645.27& -513.39\\
v=0 & -1430(10) & -1425.6 & -1419.1& -1220.0

\end{tabular}
\end{ruledtabular}
\label{tab}
\end{table}

Measured and calculated binding energies in the $0_u^+$ well are
presented in Table \ref{tab}. The measured spectrum agrees very well
with the predicted J=1 progression, except for v=1, for which the 2 MHz
discrepancy remains unexplained. Although it is too weak to be
observed, the $J=3$ progression has been calculated to illustrate the
expected rotational splitting. Finally, comparison with calculated
results for a non-retarded dipole-dipole interaction demonstrates the
significant influence of retardation. We note that our measurement
confirms the assumed value of $C_3$ to within $0.3 \%$.

In related work, we have measured other molecular lines, not belonging
to purely long-range potentials, to the red of $2^3S_1+2^3P_2$. Their
identification is under way. Additionally, an accurate measurement of
the heat deposited in the cloud by decaying long-range molecules is in
development; this could lead to a quantitative ``calorimetric" method
of measuring PA rates and thermal transport properties in the sample.
Finally, this letter provides a foundation for a two-color, stimulated
Raman experiment that would prepare molecules in the most weakly bound
state of the $^5\Sigma_g^+$ potential shown in Fig. \ref{Fig:Calculs},
with the $0_u^+$, v=0 state as an intermediate state. The corresponding
Franck-Condon factor of $\sim 0.1$ suggests that the transition rate
should be high. The lifetime of these atypical dimers formed from two
$2^3S_1$ metastable atoms is unknown. Moreover, an accurate measurement
of the s-wave elastic scattering length of $2^3S_1$ helium should
follow directly from the measured binding energy \cite{Abraham}.

{\bf Acknowledgements :} The authors thank Peter van der Straten
(Utrecht), the Cold Atoms and Molecules group at Laboratoire Aim\'e
Cotton and the Cold Atoms group at Laboratoire Kastler Brossel for
fruitful discussions. The work of APM is part of the research program
of {\it Stichting voor Fundamenteel Onderzoek der Materie}\ (FOM) which
is financially supported by {\it Nederlandse Organisatie voor
Wetenschappelijk Onderzoek}\ (NWO). MW was partly funded by National
Science Foundation Grant PHY-0140135, and TM by the {\it Procope}
exchange program.


\begin{thebibliography}{30}

\bibitem{Stwalley} W.C. Stwalley, Y.-H. Uang, G. Pichler, Phys. Rev.
Lett. {\bf 41}, 1164 (1978).

\bibitem{Lett} P. D. Lett, P.S. Julienne, W.D. Phillips, Annual Rev.
Phys. Chem. {\bf 46}, 423 (1995).

\bibitem{Cline} R.A. Cline, J.D. Miller, D.J. Heinzen, Phys. Rev. Lett.
{\bf 73}, 632 (1994).

\bibitem{Mcalexander} W.I. McAlexander, E.R.I. Abraham, R.G. Hulet, Phys. Rev. A {\bf 54},
R5 (1996).

\bibitem{Jones} K. M. Jones, P. S. Julienne, P. D. Lett, W. D.
Phillips, E. Tiesinga, C. J. Williams, Europhys. Lett. {\bf 35}, 85
(1996).

\bibitem{Wang} H. Wang, J. Li, X. T. Wang, C. J. Williams, P. L. Gould, W.
C. Stwalley, Phys. Rev. A {\bf 55}, R1569 (1997).

\bibitem{Gutterres} R.F. Gutterres, C. Amiot, A. Fioretti, C. Gabbanini,
M.Mazzoni, O. Dulieu, Phys. Rev. A {\bf 66 }, 024502 (2002).

\bibitem{Abraham} E. R. I. Abraham, W. I. McAlexander, C.A. Sackett, R. G. Hulet, Phys. Rev. Lett. {\bf 74}, 1315 (1995).

\bibitem{Gardner} J. R. Gardner, R. A. Cline, J. D. Miller, D. J.
Heinzen, H. M. J. M. Boesten, B. J. Verhaar, Phys. Rev. Lett. {\bf 74},
3764 (1995).


\bibitem{PrivateCom} P. van der Straten, private communication.


\bibitem{Herschbach} N. Herschbach, P. J. J. Tol, W. Vassen, W. Hogervorst, G. Woestenenk, J.W. Thomsen, P. van der
Straten, A. Niehaus, Phys. Rev. Lett. {\bf 84}, 1874, (2000).

\bibitem{Pereira101} F. Pereira Dos Santos, J. L\'eonard, Junmin
Wang, C. J. Barrelet, F. Perales, E. Rasel, C. S. Unnikrishnan, M.
Leduc, C. Cohen-Tannoudji, Phys. Rev. Lett. {\bf 86}, 3459 (2001).

\bibitem{Robert} A. Robert, O. Sirjean, A. Browaeys, J. Poupard, S.
Nowak, D. Boiron, C. I. Westbrook, A. Aspect, Sci. Mag. {\bf 292}, 463
(2001).

\bibitem{Pillet} P. Pillet, A. Crubellier, A. Bleton, O. Dulieu, P.
Nosbaum, I. Mourachko, F. Masnou-Seeuws, J. Phys. B {\bf 30}, 2801
(1997).

\bibitem{Dashevskaya} E.I. Dashevskaya, A.I. Voronin, E.E. Nikitin,
Can. J. Phys. {\bf 47}, 1237 (1969).

\bibitem{Meath} W. J. Meath, J. Chem. Phys. {\bf 48}, 227 (1968).

\bibitem{Drake} G. W. F. Drake in {\it Atomic, Molecular and Optical
Physics Handbook}, edited by G. W. F. Drake, AIP Press, Chap.11 (1996).

\bibitem{MovrePichler} M. Movre and G. Pichler, J. Phys. B, {\bf 10}, 2631 (1977).

\bibitem{FineStructSplit1} M.C. George, L.D. Lombardi, E.A. Hessels,
Phys. Rev. Lett. {\bf 87}, 173002, (2001), and references therein.

\bibitem{FineStructSplit2} J. Castillega, D. Livingston, A. Sanders, D.
Shiner, Phys. Rev. Lett. {\bf 84}, 4321, (2000), and references
therein.

\bibitem{Starck} J. St\"{a}rck, W. Meyer, Chem. Phys. Lett. {\bf 225},
229, (1991).

\bibitem{Hougen} J. T. Hougen, Nat. Bur. Stand. (U.S.), Monograph 115 (1970).

\bibitem{Pereira2} F. Pereira dos Santos, J. Leonard, Junmin Wang, C.
J. Barrelet, F. Perales, E. Rasel, C. S. Unnikrishnan, M. Leduc, C.
Cohen-Tannoudji, Eur. Phys. J. D, {\bf 19}, 103 (2002).

\end{thebibliography}
\end{document}